\documentclass[%
 reprint,
superscriptaddress,
 amsmath,amssymb,
 aps,
]{revtex4-2}

\usepackage{graphicx}% Include figure files
\usepackage{dcolumn}% Align table columns on decimal point
\usepackage{bm}% bold math
\usepackage{algorithm2e}

\usepackage{todonotes}

\usepackage{hyperref}% add hypertext capabilities
\usepackage{amsthm}
\usepackage{comment}
\usepackage{enumitem}
\usepackage{mathtools}
\usepackage{natbib}
\usepackage{amsfonts}
\usepackage{amssymb}

\newcommand{\mg}{\mathcal G}

\newcommand{\me}{\mathcal E}
\newcommand{\mw}{\mathcal W}
\newcommand{\mv}{\mathcal V}
\newcommand{\pp}{\mathcal P}
\newcommand{\ppq}{\mathcal Q}

\newcommand{\smin}{\setminus}
\newcommand{\nn}{\mathcal{NN}}
\newcommand{\equi}[1]{%
  #1/{\sim}%
  }
  
  \renewcommand{\vec}{\mathbf}
\newcommand{\mat}{\mathbf}

\usepackage {tikz}
\usetikzlibrary {positioning}
\usetikzlibrary{decorations.pathreplacing}

\tikzset{small dot/.style={fill=black,circle,scale=1.5}}
\tikzset{small dot 1/.style={fill=red!40,circle,scale=1.5}}
\tikzset{small dot 2/.style={fill=blue!40,circle,scale=1.5}}
\tikzset{small dot 3/.style={fill=green!40,circle,scale=1.5}}
\tikzset{small dot 4/.style={fill=red!70,circle,scale=1.5}}
\tikzset{small dot 5/.style={fill=blue!70,circle,scale=1.5}}
\tikzset{small dot 6/.style={fill=green!70,circle,scale=1.5}}

\usepackage{subcaption}
\usepackage{bbm}

\usepackage{lmodern}
\usepackage{physics}

\usetikzlibrary{matrix}

\usetikzlibrary{fit}

\usetikzlibrary{backgrounds}

\begin{document}

\title{Percolation and matrix spectrum through NIB message passing}

\author{Pedro Hack}
\email{pedro.hack@dlr.de}
\affiliation{German Aerospace Center, Germany}
\affiliation{Technical University of Munich, Germany}

\begin{abstract}
Given its computational efficiency and versatility, belief propagation is the most prominent message passing method in several applications.
In order to diminish the damaging effect of loops on its accuracy, the first explicit version of generalized belief propagation for networks, the KCN-method, was recently introduced. This approach was originally developed in the context of two target problems: percolation and the calculation of the spectra of sparse matrices. Later on, the KCN-method was extended in order to deal with inference in the context of
probabilistic graphical models on networks. It was in this scenario where an improvement on the KCN-method, the NIB-method, was conceived. We show here that this improvement can also achieved in the original applications of the KCN-method, namely percolation and matrix spectra.
\end{abstract}

\maketitle

\section{Introduction}

Message passing schemes have been shown to be key in order to address problems in several areas which are based on graphs and hypergraphs. This includes statistical mechanics, general constraint satisfaction problems, disease spread and even quantum error correction \cite{richardson2008,mezard2009,liu2019}.

Given its low time complexity, the most widely used message passing algorithm is \textbf{belief propagation} (BP). Despite its advantages, BP is know to suffer from accuracy losses when dealing with short loops. As a result, \textbf{generalized belief propagation} (GBP) \cite{yedidia2000generalized,welling2004choice} was introduced. While providing a basis for improving on BP, the main issue with generalized belief propagation was its lack of concreteness. In fact, the first explicit and general instance of GBP, the  so-called \textbf{KCN-method} \cite{hack2024belief,hack2025nib}, appeared only around two decades after the introduction of GBP \cite{cantwell2019message}.

The KCN-method was originally developed in order to target two specific problems:
\begin{itemize}
    \item percolation \cite{stauffer2018introduction,karrer2014percolation,newman2000efficient},
    \item and the computation of the spectra of sparse symmetric matrices.
\end{itemize} 
The method was then extended to inference problems in the context of probabilistic graphical models on networks and statistical mechanics \cite{kirkley2021belief}, and has since then it has found applications in a wide variety of contexts \cite{bianconi2024theory,xiong2025regulation,castro2025message,hack2024belief}. The interested reader may find the overview in \cite{newman2023message} very useful.

Recently, an improvement on the KCN-method, the so-called \textbf{NIB-method} \cite{hack2025nib}, was introduced in the context of network inference. Since no further applications of the NIB-method have been considered, our purpose here is to show that the NIB-method also provides an improvement on the KCN-method in the context of the original target problems for which the KCN-method was developed. 

\subsection{Contribution}

Our main contributions are the following:
\begin{itemize}
    \item We extend the NIB-method in order to deal with percolation (Section \ref{sec: percolation}).
    \item We extend the NIB-method in order to compute matrix spectra (Section \ref{sec: matrix spec}).
\end{itemize}

\section{Target problems}

\subsection{Percolation}

\begin{figure}
\begin{subfigure}{0.5\textwidth}
 \centering
\begin{tikzpicture}[scale=0.5, every node/.style={transform shape}]
\node[small dot] at (-2,0) (1) {};
\node[small dot] at (2,0) (2) {};
\node[small dot] at (-4,2) (3) {};
\node[small dot] at (4,2) (4) {};
\node[small dot] at (-2,4) (5) {};
\node[small dot] at (2,4) (6) {};

\path[draw,thick,-]
    (3) edge node {} (6)
    (3) edge node {} (4)
    (3) edge node {} (5)
    (5) edge node {} (4)
    (2) edge node {} (4)
    (3) edge node {} (2)
    (5) edge node {} (1)
    (1) edge node {} (6)
    (1) edge node {} (4)
    (1) edge node {} (2)
    ;
\end{tikzpicture}
  \caption{}
 % \label{fig: intersection a}
\end{subfigure}%
\\
\begin{subfigure}{0.5\textwidth}
 \centering
\begin{tikzpicture}[scale=0.5, every node/.style={transform shape}]
\node[small dot 2] at (-2,0) (1) {};
\node[small dot 1] at (2,0) (2) {};
\node[small dot 2] at (-4,2) (3) {};
\node[small dot 1] at (4,2) (4) {};
\node[small dot 2] at (-2,4) (5) {};
\node[small dot 2] at (2,4) (6) {};

\path[draw,thick,-,blue]
    (3) edge node {} (5)
    (5) edge node {} (1)
    (6) edge node {} (1)
    (3) edge node {} (6)
    ;
    
    \path[draw,thick,-,red]
    (2) edge node {} (4)
    ;
\end{tikzpicture}
  \caption{}
  %\label{fig: intersection b}
\end{subfigure}%
\caption{A (connected) base graph (a) and some realized graph associated to it (b). The realized graph consists of two connected components which are colored in red and blue, respectively.}
\label{fig: percolation graph}
\end{figure}
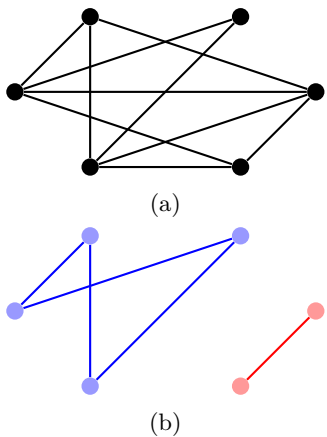

We assume here we are given some \textbf{base graph} $\mg=(\mv,\me)$ which represents the potential connections between pairs of nodes $i,j \in \mv$. We assume the base graph to be connected. However, its connections may not be actually available. In fact, we throw a coin independently for each edge $e \in \me$ such that $e$ becomes \textbf{available} or \textbf{occupied} with probability $p$ \cite{karrer2014percolation,cantwell2019message}. That is, the \textbf{realized} or \textbf{occupied graph} $\mg'$ is a subgraph of the base graph, $\mg' \subseteq \mg$. We include a base graph and some realized graph in Figure \ref{fig: percolation graph}.

Our aim is to understand the distribution of the sizes of the connected subgraphs or \textbf{clusters} and to determine the existence of a \textbf{percolating cluster}, that is, a cluster occupying a non-vanishing network fraction in the limit of large size. That is, our aim is to understand the distribution of realized graphs $\mg'$ for some fixed base graph $\mg$.

Since we are randomly occupying the available edges, we can associate to each node $i \in \mv$ a random variable $\Gamma_i \subseteq \mv \cap N_i$ which consists of the set of variables in $N_i$ which are reachable from $i$ traversing only occupied edges in some specific configuration.

Our purpose is to compute the following quantities:
\begin{itemize}
    \item The main quantity of interest is the probability that node $i$ belongs to a non-giant cluster of size $s$, $\pi_i(s)$.
    \item Given $\pi_i(s)$, we can compute the probability that node~$i$ belongs to a small cluster (of any size), and also the expected fraction of the network that belongs to the percolation cluster. Lastly, we can compute the expected size of the clusters that $i \in \mv$ belongs to, $\langle s_i \rangle \equiv \sum_{s} s \pi_i(s)$.
\end{itemize}

 We explain how to deal with percolation through the NIB-method in Section \ref{sec: percolation}.

\subsection{Matrix spectra}

We assume here we are given some $n \times n$ symmetric matrix $\mat{A}$. Our aim is to compute the \textbf{spectrum} of $\mat{A}$, that is, its set of eigenvalues. In order to do so, we can approximate its \textbf{spectral density}
\begin{equation}
\rho(x) \equiv \frac{1}{n} \sum_{i=1}^{n} \delta(x - \lambda_i),
\label{eq:density1}
\end{equation}
where $\{\lambda_i\}_i$ are the eigenvalues of $\mat{A}$, and $\delta(\cdot)$ is Dirac's delta.

By Using \cite[Eq. 21]{nadakuditi2013spectra}, and taking and $z \equiv x + i\eta$, one can show~\cite{cantwell2019message} that the spectral density \eqref{eq:density1} is approximately the imaginary part of the complex function
\begin{align}
\rho(z) \equiv -\frac{1}{n \pi} \sum_{i=1}^{n} \frac{1}{z-\lambda_i} = -\frac{1}{n \pi z} \sum_{s=0}^{\infty}\sum_{i=1}^{n} \frac{X_i^s}{z^s},
\label{eq:density2}
\end{align}
where $X_i^s \equiv [\mat{A}^s]_{ii}$ is the $i$th diagonal element of~$\mat{A}^s$. In order for \eqref{eq:density2} to
accurately approximate \eqref{eq:density1}, one ought to take the limit as the imaginary part $\eta\to0$ from above. In fact, $\eta$ is a resolution parameter that broadens the peaks in~\eqref{eq:density1} by approximately its value \footnote{More specifically, given some $x \in \mathbb R$ of interest, we use $z= x + i \eta_0$ for some fixed $\eta_0>0$. For instance, in \cite{cantwell2019message}, $\eta_0=0.05, 0.01$. When using the message passing methods that we will present later on, one runs them with such a fixed value and, when convergence is achieved, we simply take the imaginary part of $\rho(z)$. Hence, we will run the algorithm once for each value of $x$.}.

We can associate to every $n \times n$ symmetric matrix $\mat{A}$ a \textbf{weighted graph} $\mg_{\mat{A}}=(\mv, \me, \mw)$, where
\begin{itemize}
    \item we associate one vertex to each index that a column or row of $\mat{A}$ may take,  $\mv \equiv \{1,\dots,n\}$;
    \item given $i,j \in \mv$ with $i \leq j$, then $(i,j) \in \me$ if and only if $[\mat{A}]_{ij} \neq 0$ \footnote{We could avoid this condition and simply take $\mg_A$ to have full connectivity with some weights being null. However, since our message passing methods will be exploiting the sparsity of $\mat{A}$, it is more convenient to associate a sparse graph $\mg_{\mat{A}}$ to a sparse matrix $\mat{A}$.};
    \item if $(i,j) \in \me$, then $w_{(i,j)}= [\mat{A}]_{ij}$.
\end{itemize}

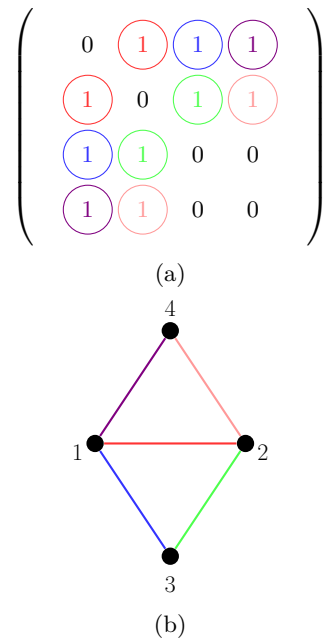
\begin{figure}
\begin{subfigure}{0.5\textwidth}
 \centering
\begin{tikzpicture}[every node/.style={minimum size=2.25em}] \matrix (m) [matrix of math nodes, nodes in empty cells,
  left delimiter={(}, right delimiter={)}] at (0,0)
  {
     0 &  |[draw,red!80,circle,inner sep=0pt,minimum size=2em]| 1 & |[draw,blue!80,circle,inner sep=0pt,minimum size=2em]| 1 & |[draw,violet,circle,inner sep=0pt,minimum size=2em]| 1 \\
  |[draw,red!80,circle,inner sep=0pt,minimum size=2em]| 1 & 0 & |[draw,green!70,circle,inner sep=0pt,minimum size=2em]| 1 & |[draw,red!40,circle,inner sep=0pt,minimum size=2em]| 1 \\
  |[draw,blue!80,circle,inner sep=0pt,minimum size=2em]| 1 & |[draw,green!70,circle,inner sep=0pt,minimum size=2em]| 1 & 0 & 0 \\
   |[draw,violet,circle,inner sep=0pt,minimum size=2em]| 1 & |[draw,red!40,circle,inner sep=0pt,minimum size=2em]| 1 & 0 &  0 \\
  };
\end{tikzpicture}
  \caption{}
  \label{fig: intersection a}
\end{subfigure}%
\\
\begin{subfigure}{0.5\textwidth}
  \centering
  \begin{tikzpicture}[scale=0.5, every node/.style={transform shape}]
\node[small dot, label={[anchor=north,above=.5cm, thick, font=\fontsize{18}{18}\selectfont]270:\textbf{$4$}}] at (2,6) (TU) {};

\node[small dot, label={[anchor=east,right=.2cm, thick, font=\fontsize{18}{18}\selectfont]270:\textbf{$2$}}] at (4,3) (TM2) {};
\node[small dot, label={[anchor=west,left=.2cm, thick, font=\fontsize{18}{18}\selectfont]270:\textbf{$1$}}] at (0,3) (TM1) {};

\node[small dot, label={[anchor=south,below=.2cm, thick, font=\fontsize{18}{18}\selectfont]270:\textbf{$3$}}] at (2,0) (TD) {};

\path[draw,thick,-,violet]
(TM1) edge node {} (TU)
 ;
\path[draw,thick,-,red!40]
(TM2) edge node {} (TU)
 ;
\path[draw,thick,-,blue!80]
(TD) edge node {} (TM1)
 ;
\path[draw,thick,-,green!70]
(TD) edge node {} (TM2)
 ;
\path[draw,thick,-,red!80]
(TM1) edge node {} (TM2)
    ;
    
\end{tikzpicture}
  \caption{}
  \label{fig: intersection b}
\end{subfigure}%
\caption{A symmetric matrix (a) and its associated weighted graph (b). We color the entries of the matrix in the same color as the edges in the associated graph. Since the weights are all equal in this example, we do not include them in the figure.}
\label{fig: spectum graph}
\end{figure}

We include an example of the weighted graph associated to some symmetric matrix in Figure \ref{fig: spectum graph}.

Returning to Section \ref{sec: single rv factors}, note that $\mg_{\mat{A}}$ will have self-loops $(k,k) \in \me$ whenever we have some non-zero diagonal term $A_{kk} \neq 0$. It will become clear later on that we cannot absorb these self-loops into pairwise potentials and, hence, we must consider the variation of the NIB-method proposed in Section \ref{sec: single rv factors}. 

A \textbf{closed walk} that starts and ends at~$i \in \mv$, or \textbf{$i$-walk} to be precise, is a sequence of vertices $(i_0,\dots,i_m)$ such that $i_0=i_m=i$ and $(i_t,i_{t+1}) \in \me$ for $0 \leq t \leq m-1$. Such a walk has an associated \textbf{weight}, which is the product of the matrix elements on the edges traversed by the walk. Closed walks and their weights are important in order to study matrix spectra since 
$X_i^s$ in \eqref{eq:density2} equals the sum of the weights of all the length $s$ closed walks on $\mg_{\mat{A}}$ that start and end at~$i \in \mv$.

An \textbf{excursion}, or an \textbf{$i$-excursion} to be more precise, is a closed walk that starts in $i \in \mv$ and only returns once to $i$ (i.e. at the \textbf{end} of the walk). This means that any closed walk $c$ returning $m$ times to $i$ can be decomposed as a succession of $m$ excursions $(w_i)_{i=1}^m$, with the length of $c$ being $s$ provided $s=\sum_{u=1}^m s_u$, where $s_i$ is the length of $w_i$ for $i=1,\dots,m$.

Excursions are key in our study of matrix spectra through \eqref{eq:density2}. This is the case since, if we denote by $Y_i^s$ the sum of the weights of all excursions of length~$s$
that start and end at node~$i$, then, given that
\begin{equation}
X_i^s = \sum_{m=0}^{\infty} \left[ \sum_{s_1=1}^{\infty} \dots \sum_{s_m=1}^{\infty}  \delta \bigl(s, {\textstyle\sum_{u=1}^{m}} s_u \bigr) \prod_{u=1}^{m} Y_i^{s_u} \right],
\label{eq:X_i^s}
\end{equation}
we get that \eqref{eq:density2} becomes
\begin{equation}
\rho(z) = -\frac{1}{n \pi z} \sum_{i=1}^{n} \sum_{m=0}^{\infty}\,\prod_{u=1}^{m} \Biggl[ \sum_{s=1}^{\infty} \frac{ Y_i^s }{z^{s}} \Biggr].
\label{eq:rhoy}
\end{equation}
Moreover, defining 
\begin{equation}
H_i(z) \equiv \sum_{s=1}^{\infty} \frac{Y_i^s}{z^{s-1}},
\label{eq:closed_walk_gen_fun}
\end{equation}
we obtain
\begin{equation}
\rho(z) = -\frac{1}{n \pi z} \sum_{i=1}^{n} \sum_{m=0}^{\infty} \biggl[ \frac{H_i(z)}{z} \biggr]^m 
  = -\frac{1}{n \pi} \sum_{i=1}^{n} \frac{1}{z-H_i(z)}.
\label{eq:final_rho}
\end{equation}
Thus, we can compute~$\rho(z)$ and determine the spectrum of $\mat{A}$ through~$H_i(z)$. We explain how to compute $H_i(z)$ through the NIB-method in Section \ref{sec: matrix spec}.

\section{The NIB-method}

The NIB-method \cite{hack2025nib} was introduced in the context of (network) graphical model inference, where one starts with a connected graph $\mg = (\mv,\me)$ that is associated with a probability distribution $p$, that is, where the nodes $i \in \mv$ represent random variables $X_i$ over which $p$ is defined and the edges $\me$ represent a factorization of $p$ in terms of pairs of random variables $\{X_i,X_j\}$
\begin{equation}
 \label{eq: distribution}
     p(x_1,\dots,x_{|\mv|}) \propto \prod_{(i,j) \in \mg} f_{i,j}(x_i,x_j),
 \end{equation}
 where $f_{i,j}: X_i \times X_j \to \mathbb R_{\geq 0}$ and we omit a normalization constant. We assume for the moment that there are no \textbf{self-loops}, that is, $(k,k) \not \in \mg$ for all $k \in \mv$. We return to this point in Section \ref{sec: single rv factors}.
 
At its core, the NIB-method is a procedure that, given a problem that can be encoded as a graph, allows one to infer properties of interest related to that graph via a message-passing scheme. Since
message-passing schemes are known to suffer from accuracy loss when facing loops, the NIB-method provides an explicit recipe to break $\mg$ into subgraphs such that one can improve the accuracy of the standard message-passing schemes like BP by exchanging messages between these subgraphs.

Formally, the NIB-method uses an underlying integer parameter $r \geq 0$, the \textbf{loop bound}, and defines a message-passing scheme for each value of $r$. In order to do so, it considers two types of neighborhoods:
\begin{itemize}
    \item the \textbf{primary} neighborhoods
    \begin{equation}
     \label{eq: prim neigh}
         \{N_i^{(r)}\}_{i \in \mv},
     \end{equation}
     where $N_i^{(r)}$ consists of $i$ together with its nearest neighbors $\nn_i$ and the edges joining it to them, plus both edges and nodes along paths that join two nearest neighbors of $i$;
    \item and the \textbf{intersection} neighborhoods
 \begin{equation}
 \label{eq: neigh inter}
 \begin{split}
     &\{N_{i \cap j}^{(r)} \}_{i \in \mv, j \in N_i^{(r)} \smin \{i \}}, \text{ where} \\
     &N_{i \cap j}^{(r)} \equiv N_i^{(r)} \cap N_j^{(r)}
     \end{split}
 \end{equation}
 and $\cap$ is the usual set intersection.
\end{itemize}

Another important set of neighborhoods, which are not used in the NIB-method by appear in the KCN-method, are
\begin{itemize}
    \item the \textbf{difference} neighborhoods 
    \begin{equation}
     \label{eq: diff neigh}
     \{N_{i \smin j}^{(r)}\}_{i \in \mv, j \in N_i^{(r)} \smin \{i\}},    
     \end{equation}
     where $N_{i \smin j}^{(r)}$ consists of node $i$ together with all the edges in $N_i^{(r)}$ which are not in $N_j^{(r)}$, and the nodes at the endpoints of such edges.
\end{itemize}
In the following, we drop the superscript $(r)$ for commodity.

Depending on whether the loop bound $r$ is \textbf{fulfilled}, i.e. all loops around $i$ are contained within $N_i^{(r)}$ for each $i \in \mv$, we distinguish two instances of the NIB-method: \textbf{$r$-bounded loops} (if it is fulfilled), and \textbf{$r$-unbounded loops} (if it is not). As remarked in \cite{hack2025nib}, both cases can be introduced together, although distinguishing between them is useful from a pedagogical point of view.

\subsection{$r$-bounded loops}

If the loop bound is \textbf{fulfilled}, then one can associate to $\mg$ a \textbf{hypernetwork} $\equi{\mg}$ that is loopless. We can do so because of the
\textbf{equivalence class condition}, which states that, for $i \in \mv$ and $j \in \mv \cap \left( N_i \smin \{i \} \right)$, we have
\begin{equation}
\label{eq: equiv class cond}
    N_{i \cap j} = N_{k \cap q}
\end{equation}
 for all $k,q \in \mv \cap N_{i \cap j}$, $k \neq q$. Taking this into account, we introduce an equivalence relation $\sim$ on the intersection neighborhoods,
 \begin{equation*}
 N_{i \cap j} \sim N_{k \cap q} \text{ if and only if } N_{i \cap j} = N_{k \cap q},
 \end{equation*}
 and end up with a hypernetwork
  \begin{equation*}
 %\begin{split}
     \equi{\mg} \equiv \left(\equi{\cap}, \{ e_s\}_{s \in \text{Piv}(\mg)} \right)
% \end{split}
\end{equation*}
that consists of the equivalence classes as nodes
 \begin{equation*}
\equi{\cap} \equiv \left\{ \overline{i \cap j} \right\}_{i \in \mv, j \in N_i \smin \{i\}},
\end{equation*}
with $\overline{i \cap j}$ being the equivalence class of $N_{i \cap j}$, and one hyperedge 
 \begin{equation*}
e_s \equiv \{ \overline{i \cap j} \in \equi{\cap} | s \in \overline{i \cap j} \}
\end{equation*}
for each node in the \textbf{pivots set} $s \in \text{Piv}(\mg)$, that is, for each $s \in \mv$ that belongs to at least two different equivalence classes in $\equi{\cap}$.

Since the loop bound is fulfilled, $\equi{\mg}$ is loopless and the NIB-method provides exact results by exchanging messages between the equivalence classes through the hyperedges
\begin{equation}
\label{eq: messages nib}
        \{m^{(t)}_{\overline{i \cap j} \to i}\}_{i \in V, j \in (N_i \smin \{i\})/\sim, t \geq 0}
\end{equation}
where, given $j,k \in N_i \smin \{i\}$,
\begin{equation*}
    j \sim k \text{ if and only if } \overline{i \cap j} = \overline{i \cap k}.
\end{equation*}
The exact form of the messages in the context of (network) graphical model inference \cite{hack2025nib} is not relevant for our purposes here.

\subsection{$r$-unbounded loops}
\label{sec: nib unbounded}

If the loop bound is \textbf{not} fulfilled, then one cannot associate to $\mg$ a loopless hypernetwork anymore. Moreover, the neighborhood intersections do not constitute equivalence classes and may overlap in non-trivial ways. Thus, if we still want to use the neighborhoods intersections in this context and in order to avoid unnecessary errors, we ought to find ways of coping with overcounting during the message update and inference phases. In order to do so, and taking $2^\me$ to be the power set of $\me$, \cite{hack2025nib} introduces two maps:
\begin{itemize}
    \item $\overline{\pp_{i \cap j}}: \{N_{k \cap q}\}_{k \in N_{i \cap j}, q \in N_k} \to 2^\me$.
To define $\overline{\pp_{i \cap j}}(\cdot)$, we first recursively define the set $\pp_{i \cap j}$ as follows:
\begin{itemize}
    \item we initialize it by including all the functions within $N_{i \cap j}$;
    \item at each following step, we pick some $k \in N_{i \cap j}$ and some $q \in N_k$ and we incorporate the functions within $N_{k \cap q}$ to $\pp_{i \cap j}$.
\end{itemize}
Lastly, we take $\overline{\pp_{i \cap j}}(N_{k \cap q})$ to be $\pp_{i \cap j}$ at the step right before the pair $k,q \in \mv$ is selected.
\item $\overline{\ppq_{i}}: \{N_{i \cap j}\}_{j \in N_{i} \smin \{i\}} \to 2^\me$. To define $\overline{\ppq_{i}}(\cdot)$, we first recursively define the set $\ppq_{i}$ as follows:
    \begin{itemize}
        \item we initialize it as the empty set;
        \item at each following step, we pick some $j \in N_{i} \smin \{ i\}$ and we incorporate the functions within $N_{i \cap j}$ to $\ppq_{i}$.
    \end{itemize}
Lastly, we take $\overline{\ppq_{i}}(N_{i \cap j})$ to be $\ppq_{i}$ at the step right before $j \in \mv$ is selected.
\end{itemize}

Given some intersection $N_{i \cap j}$, the messages it receives in this case
 \begin{equation*}
        \{m^{(t)}_{k \cap q \to i \cap j}\}_{i \in V, j \in N_i \smin \{i\}, k \in N_{i \cap j}, q \in  N_k,t \geq 0},
\end{equation*}
are sent from all the intersections $N_{k \cap q}$, such that  $k \in N_{i \cap j}$ and $q \in N_k$. The exact form of the messages in the context of (network) graphical model inference \cite{hack2025nib} is again not relevant for our purposes here.

\subsection{Single variable factors and self-loops}
\label{sec: single rv factors}

In its original form, the NIB-method assumes we are given some probability distribution $p$ of the form \eqref{eq: distribution}, and it emphasizes that, in case we also had some single-variable functions $f_{k}: X_k \to \mathbb R_{\geq 0}$ or \textbf{external potentials} in the product decomposition of $p$,
\begin{equation*}
     p(x_1,\dots,x_{|\mv|}) \propto \prod_{(i,j) \in \mg} f_{i,j}(x_i,x_j) \prod_{k \in \mg} f_{k}(x_k),
 \end{equation*}
then we could absorb $(f_k)_{k \in \mg}$ into the pairwise potentials $( f_{i,j} )_{(i,j) \in \mg}$.

Alternatively, we could incorporate to the NIB-method the \textbf{trivial intersection neighborhoods} 
\begin{equation*}
    N_{k \cap k} \equiv \{k,(k,k)\},
\end{equation*}
where $(k,k)$ is the \textbf{self-loop} in $\mg$ associated to $f_k$, which conform \textbf{trivial equivalence classes} $\overline{k \cap k}$ that we add to $\equi{\cap}$ in the $r$-bounded case. Regarding message-passing, this amounts to adding some \textbf{trivial messages}, that is, messages from the trivial intersection neighborhood $N_{k \cap k}$ to node $k$. This messages provide the information in $f_k$ to $k$ and take the same form in both the bounded $m^{(t)}_{\overline{k \cap k} \to k}$ and unbounded $m^{(t)}_{k \cap k \to i \cap j}$ cases, and are incorporated in each update and inference equation that $k$ participates in. 

Although the distinction between these two versions of the NIB-method are not important in the context of graphical model inference, they can be meaningful in some applications. In fact, although we can disregard them when studying percolation, they \textbf{must} be considered when computing matrix spectra, that is, we cannot incorporate them into pairwise interactions as in the original NIB-method in the latter case. We will return to this in Section \ref{sec: matrix spec}.

\section{Percolation via the NIB-method}
\label{sec: percolation}

We explain how to deal with percolation through the NIB-method in this section, distinguishing the cases where the base graph $\mg$ fulfills the loop bound (Section \ref{sec: percolation bounded}) from those where it does not (Section \ref{sec: percolation unbounded}).
Regarding the discussion in Section \ref{sec: single rv factors}, note that self-loops are superfluous for our purposes, since they cannot connect different nodes and hence they do not affect cluster sizes.

\subsection{r-bounded loops}
\label{sec: percolation bounded}

In order to compute $\pi_i(s)$, we first compute $\pi_i(s|\Gamma_i)$, the probability that $i \in \mv$ belongs to a cluster of size $s$ given some configuration of occupied edges in $N_i$, $\Gamma_i$.

Given some $\Gamma_i$ and some $j \in \Gamma_i$, and taking as $s_{\overline{j \cap k}}$ the size of the cluster that node $j$ would belong to provided we remove  from $\mg$ all the equivalence classes that $j$ belongs to except for $\overline{j \cap k}$, we get that 
\begin{widetext}
\begin{equation}
\label{eq: condi prob cluster}
\pi_i(s| \Gamma_i) = \!\!\sum_{\{ s_{\overline{j \cap k}} : j \in \Gamma_i \smin \{i\}, k \in \equi{(N_j \smin \{i,j\})}\}}
   \biggl[ \prod_{j \in \Gamma_i \smin \{i\}} \prod_{k \in \equi{(N_j \smin \{i,j\})}} \pi_{\overline{j \cap k} \to j}(s_{\overline{j \cap k}}) \biggr] \delta(s-1, \textstyle\sum_{j \in \Gamma_i \smin \{i\}, k \in \equi{(N_j \smin \{i,j\})}} s_{\overline{j \cap k}}),
\end{equation}
\end{widetext}
where $\pi_{\overline{j \cap k} \to j}(s)$ is the probability that node~$j$ is in a cluster of size~$s$ once the edges in the equivalence classes $\overline{j \cap q} \neq \overline{j \cap k}$ have been removed, and $\delta(\cdot,\cdot)$ is the Kronecker delta. The condition on the delta comes from the fact that a cluster to which $i$ belongs should have the same size (minus one since we remove precisely $i$) as the sum of the sizes $s_{\overline{j \cap k}}$ of the clusters that each elements in $N_i$ which is connected to $i$ via $\Gamma_i$ belongs to once we have removed the edges in $\overline{j \cap q} \neq \overline{j \cap k}$.

For our purposes, it is useful \cite{newman2000efficient,cantwell2019message} to define a generating function for $\pi_i(s|\Gamma_i)$:
\begin{equation*}
    H_i(z|\Gamma_i) \equiv \sum_{s} \pi_i(s|\Gamma_i)\,z^s.
\end{equation*}
In fact, in our setup, we have that
\begin{widetext}
\begin{align}
H_i&(z|\Gamma_i) =\sum_{s} z^{s} \!\!\sum_{\{ s_{\overline{j \cap k}} : j \in \Gamma_i \smin \{i\}, k \in \equi{(N_j \smin \{i,j\})} \}}
   \biggl[ \prod_{j \in \Gamma_i \smin \{i\}} \prod_{k \in \equi{(N_j \smin \{i,j\})}} \pi_{\overline{j \cap k} \to j}(s_{\overline{j \cap k}}) \biggr] \delta(s-1, \textstyle\sum_{j \in \Gamma_i \smin \{i\}, k \in \equi{(N_j \smin \{i,j\})}} s_{\overline{j \cap k}}) \nonumber \\
   = &z \prod_{j \in \Gamma_i \smin \{i\}} \prod_{k \in \equi{(N_j \smin \{i,j\})}} \sum_{s_{\overline{j \cap k}}} z^{s_{\overline{j \cap k}}} \pi_{\overline{j \cap k} \to j}(s_{\overline{j \cap k}}) = z \prod_{j \in \Gamma_i \smin \{i\}} \prod_{k \in \equi{(N_j \smin \{i,j\})}} H_{\overline{j \cap k} \to j}(z) \nonumber \\
    = &z \prod_{j \in \equi{(N_i \smin \{i\})}} \prod_{k \in \Gamma_{i \cap j} \smin \{i\}} \prod_{q \in \equi{(N_k \smin \{i,j\})}} H_{\overline{k \cap q} \to k}(z) = z \prod_{j \in \equi{(N_i \smin \{i\})}} \prod_{k \in \overline{i \cap j} \smin \{i\}} \prod_{q \in \equi{(N_k \smin \{i,j\})}} \left[ H_{\overline{k \cap q} \to k}(z) \right]^{w^{\overline{i \cap j}}_{ik}} \nonumber \\
    = &z \prod_{j \in \equi{(N_i \smin \{i\})}} \prod_{k \in \overline{i \cap j} \smin \{i\}} \left[ H_{\neg (\overline{i \cap j}) \to k}(z) \right]^{w^{\overline{i \cap j}}_{ik}},
\label{eq:cluster_gen_fn}
\end{align}
\end{widetext}
where we have used \eqref{eq: condi prob cluster}, we have introduced both
\begin{equation*}
\Gamma_{i \cap j} \equiv \Gamma_i \cap (\overline{i \cap j})    
\end{equation*}
and the random variable
\begin{equation}
\label{eq: rv w}
w^{\overline{i \cap j}}_{ik} \equiv \biggl\lbrace\begin{array}{ll}
	1 & \quad\mbox{if $k \in\Gamma_{i \cap j}$,} \\
  0 & \quad\mbox{otherwise,}
\end{array}
\end{equation}
that is, $w^{\overline{i \cap j}}_{ik}=1$ if there is a path of occupied edges within $\overline{i \cap j}$ from $i$ to~$k$, and, given some $k \in \overline{i \cap j}$, we use the scalar
\begin{equation}
\label{eq: not int message}
    H_{\neg (\overline{i \cap j}) \to k}(z) \equiv \prod_{q \in \equi{(N_k \smin \{i,j\})}}\!H_{\overline{k \cap q} \to k}(z).
\end{equation}

In order to compute $\pi_i(s)$, we ought to 
average $\pi_i(s|\Gamma_i)$ over the possible configurations 
$\Gamma_i$, that is, we have that $\pi_i(s) = \bigl\langle \pi_i(s|\Gamma_i) \bigr\rangle_{\Gamma_i}$, where the average is weighted via the probability of each realization $\Gamma_i$:
$p^k (1-p)^{m-k}$, where $m \equiv |N_i \cap \me|$ is the number of edges in $N_i$, and $k$ is the number of occupied edges in $N_i$.

Averaging over $\Gamma_i$ in \eqref{eq:cluster_gen_fn} we obtain
\begin{align}
H_i(z) &\equiv \sum_s \pi_i(s)\,z^s
        = \bigl\langle H_i(z|\Gamma_i) \bigr\rangle_{\Gamma_i} \nonumber \\
        &= z \prod_{j \in \equi{(N_i \smin \{i\})}} \Bigl\langle\! \prod_{k \in \overline{i \cap j} \smin \{i\}} \left[ H_{\neg (\overline{i \cap j}) \to k}(z) \right]^{w^{\overline{i \cap j}}_{ik}} \Bigr\rangle_{\Gamma_{i \cap j}} \nonumber \\
       &= z \prod_{j \in \equi{(N_i \smin \{i\})}} G_{\overline{i \cap j} \to i}\bigl( \vec{H}_{\overline{i \cap j} \to i}(z) \bigr) \nonumber \\
       &= z G_i(\vec{H}_{\to i}(z)),
\label{eq:cluster_gen_fn2}
\end{align}
where we denote by $G_{\overline{i \cap j} \to i}(\vec{y})$ a generating function for~$w^{\overline{i \cap j}}_{ik}$ in \eqref{eq: rv w},
\begin{equation*}
G_{\overline{i \cap j} \to i}(\vec{y}) \equiv  \Bigl\langle\!\prod_{k \in \overline{i \cap j} \smin \{i\}} y_k^{w^{\overline{i \cap j}}_{ik}} \Bigr\rangle_{\Gamma_{i \cap j}},
\end{equation*}
and we take
\begin{equation*}
   G_i(\vec{y}) \equiv  \prod_{j \in \equi{(N_i \smin \{i\})}} G_{\overline{i \cap j} \to i}\bigl( y_j \bigr), \end{equation*}
   and the vector of scalars in \eqref{eq: not int message} for the different 
$k$ in~$\overline{i \cap j} \smin \{i\}$
   \begin{equation*}
    \vec{H}_{\overline{i \cap j} \to i}(z) \equiv \left( H_{\neg (\overline{i \cap j}) \to k}(z) \right)_{k \in \overline{i \cap j} \smin \{i\}}.
\end{equation*}
 Lastly, we consider the concatenation of the vectors $\vec{H}_{\overline{i \cap j} \to i}(z)$ over the different equivalence classes that $i$ belongs to
\begin{equation}
    \vec{H}_{\to i}(z) \equiv \left( \vec{H}_{\overline{i \cap j} \to i}(z) \right)_{j \in \equi{(N_i \smin \{i\})}}.
\end{equation}

To conclude our calculation, we ought to evaluate the~$H_{\overline{k \cap q} \to k}(z)$. This can be done following the idea in the computation of~$H_i(z)$, the only difference being that we only consider the product over the elements in the equivalence class $\overline{k \cap q}$. That is,  we can derive a generating function
\begin{equation}
H_{\overline{k \cap q} \to k}(z| \Gamma_{k \cap q}) = z\!\!\prod_{s \in \overline{k \cap q} \smin \{k\}}  \left[H_{\neg (\overline{k \cap q}) \to s}(z) \right]^{w^{\overline{k \cap q}}_{k s}}
\end{equation}
that, once averaged over~$\Gamma_{k \cap q}$, yields
\begin{equation}
H_{\overline{k \cap q} \to k}(z) = z G_{\overline{k \cap q} \to k}\bigl( \vec{H}_{\overline{k \cap q} \to k}(z) \bigr).
\label{eq:self_consistent_message_eq}
\end{equation}
 We can solve \eqref{eq:self_consistent_message_eq} iteratively by message passing, starting with some initial random values and iterating the equations to convergence.
 We can then substitute the solution into Eq.~\eqref{eq:cluster_gen_fn2} and obtain the cluster size generating function.

From the cluster size generating function \eqref{eq:cluster_gen_fn2} we can derive other quantities of interest:
\begin{itemize}
    \item The probability that node~$i$ belongs to a small cluster of any size is $H_i(1) = \sum_s \pi_i(s)$.
    \item The expected fraction~$S$ of the network taken up by the percolating cluster is
\begin{equation}
S = 1 - \frac{1}{n} \sum_i H_i(1).
\end{equation}
This is the case since, if it does not belong to a small cluster, then a node must be in the percolating cluster.
\item The expected size fo the clusters that $i \in \mv$ belongs to is
\begin{widetext}
\begin{equation*}
%\begin{split}
    \langle s_i \rangle =
 H_i(1) + \sum_{j \in \equi{(N_i \smin \{i\})}} \sum_{k \in \overline{i \cap j} \smin \{i\}} \sum_{q \in \equi{(N_k \smin \{i,j\})}} \partial_{\overline{i \cap j}} G_{i}\bigl( \vec{H}_{ \to i}(1) \bigr) \partial_k G_{\overline{i \cap j} \to i}\bigl( \vec{H}_{\overline{i \cap j} \to i}(1) \bigr)  \partial_{\overline{k \cap q}} H_{\neg (\overline{i \cap j}) \to k}(1) H'_{\overline{k \cap q} \to k}(1),
\end{equation*}
\end{widetext}
where $H'$ is the derivative of~$H$, $\partial_{\overline{i \cap j}} G_i$ is the partial derivative of $G_i$ with respect to its $j$th argument, and the same holds for $\partial_k G_{\overline{i \cap j} \to i}$ and $\partial_{\overline{k \cap q}} H_{\neg (\overline{i \cap j}) \to k}$.

$H'_{\overline{k \cap q} \to k}(1)$ can be found by differentiating Eq.~\eqref{eq:self_consistent_message_eq}, setting~$z=1$, and iterating the self-consistent equations
\begin{equation}
 \label{eq:H_deriv}
\begin{split}
&H'_{\overline{k \cap q} \to k}(1) = \sum_{s \in \overline{k \cap q} \smin \{k\}} \sum_{v \in \equi{(N_s \smin \{k,q\})}} \partial_{s} G_{\overline{k \cap q} \to k}\bigl( \vec{H}_{\overline{k \cap q} \to k}(1) \bigr)\\
 &\times \partial_{\overline{s \cap v}} H_{\smin (\overline{k \cap q}) \to s}(1) H'_{\overline{s \cap v} \to s}(1) +H_{\overline{k \cap q} \to k}(1)
 \end{split}
\end{equation}
until convergence.
\end{itemize}

Since the loop bound is fulfilled, the equations in this section provide exact results. Moreover, they provide an advantage regarding time complexity compared to the direct application of the KCN-method to percolation \cite{cantwell2019message}: Instead of summing over $N_i$ and $N_{j \smin i}$, we only sum over $N_{i \cap j}$. In fact, following \cite[Claim 4]{hack2025nib}, we can show that the approach to percolation in this section is optimal in terms of time complexity.

\subsection{r-unbounded loops}
\label{sec: percolation unbounded}

If the loop bound is not fulfilled, then we can use the maps defined in Section \ref{sec: nib unbounded} to extend our approach to percolation from the bounded case in the spirit of the extension of the NIB-method to the unbounded case \cite{hack2025nib}.

Regarding the message passing equations, we use
\begin{equation*}
H_{k \cap q \to i \cap j}(z) \equiv z G^{\overline{\pp_{i \cap j}}(N_{k \cap q})}_{k \cap q\to i \cap j}\bigl( \vec{H}_{k \cap q \to i \cap j}(z) \bigr),
\end{equation*}
where 
\begin{equation*}
    G^{\overline{\pp_{i \cap j}}(N_{k \cap q})}_{k \cap q\to i \cap j}(y) \equiv  \Bigl\langle\!\prod_{s \in N_{k \cap q}} y_s^{w^{\overline{\pp_{i \cap j}}(N_{k \cap q})}_{ks}} \Bigr\rangle_{\Gamma_{k \cap q}}
\end{equation*}
and we have introduced the random variable~$w^{\overline{\pp_{i \cap j}}}_{ks}$ which takes the value~1 if there is a path of occupied edges within $\overline{\pp_{i \cap j}}(N_{k \cap q})$ from $k$ to~$s$, and zero otherwise. Moreover, we use the notation
\begin{equation*}
\begin{split}
  &H_{p \to k \cap q} (z) \equiv  \prod_{s \in N_{p} \smin \{p\}} \!H_{p \cap s \to k \cap q}(z) \text{, and} \\
  &\vec{H}_{k \cap q \to i \cap j}(z) \equiv \left( H_{p \to k \cap q} (z)\right)_{p \in N_{k \cap q} \smin \{k\}}.
  \end{split}
\end{equation*}

For the inference stage, we use the equation
\begin{align*}
H_i(z) &\equiv z \prod_{j \in N_i \smin \{i\}} G_{i \cap j \to i}^{\overline{\ppq_i}(N_{i \cap j})}\bigl( \vec{H}_{i \cap j \to i}(z) \bigr) \nonumber \\
       &= z G_i^{\overline{\ppq_i}}(\vec{H}_{\to i}(z)),
\end{align*}
where
\begin{equation}
\begin{split}
&G_{i \cap j \to i}^{\overline{\ppq_i}(N_{i \cap j})}(\vec{y}) \equiv  \Bigl\langle\!\prod_{k \in N_{i \cap j} \smin \{i\}} y_k^{w^{\overline{\ppq_i}(N_{i \cap j})}_{ik}} \Bigr\rangle_{\Gamma_{i \cap j}} \text{, and}\\
&G_i^{\overline{\ppq_i}}(\vec{y}) \equiv  \prod_{j \in N_i \smin \{i\}} G_{i \cap j \to i}^{\overline{\ppq_i}(N_{i \cap j})}\bigl( y_j \bigr).
\end{split}
\end{equation}
Moreover, we use the notation
\begin{equation*}
\begin{split}
    &\vec{H}_{i \cap j \to i}(z) \equiv \left( H_{k \to i \cap j} (z) \right)_{k \in N_{i \cap j} \smin \{i\}} \text{, and}\\
    &\vec{H}_{\to i}(z) \equiv \left(\vec{H}_{i \cap j \to i}(z) \right)_{j \in N_i \smin \{i\}}
    \end{split}
\end{equation*}

To conclude this section, we only ought to show how to compute the expected value of $s_i$. We can do so using the following equation
\begin{equation*}
    \begin{split}
        &\langle s_i \rangle \equiv H_i(1) + \sum_{j \in N_i \smin \{i\}} \sum_{k \in N_{i \cap j} \smin \{i\}} \sum_{q \in N_{k}} \partial_{i \cap j} G_{i}^{\overline{\ppq_i}}\bigl( \vec{H}_{ \to i}(1) \bigr)\\
        &\times \partial_k G_{i \cap j \to i}^{\overline{\ppq_i}(N_{i \cap j})}\bigl( \vec{H}_{i \cap j \to i}(1) \bigr) \partial_{k \cap q} H_{k \to i \cap j}(1) H'_{k \cap q \to i \cap j}(1),
    \end{split}
\end{equation*}
where $H'_{k \cap q \to i \cap j}(1)$ can be found by iterating the self-consistent equations
\begin{equation*}
    \begin{split}
        &H'_{k \cap q \to i \cap j}(1) \equiv \sum_{s \in k \cap q \smin \{k\}} \sum_{v \in N_s} \partial_{s} G_{k \cap q \to i \cap j}^{\overline{\pp_{i \cap j}}(N_{k \cap q})}\bigl( \vec{H}_{k \cap q \to i \cap j}(1) \bigr) \\
        &\times \partial_{s \cap v} H_{s \to k \cap q}(1) H'_{s \cap v \to k \cap q}(1)
        + H_{k \cap q \to i \cap j}(1)
    \end{split}
\end{equation*}

Since the loop bound is not fulfilled, these equations only provide approximate results. The time complexity advantage compared to the KCN-method remains, and the accuracy does not decrease provided we consider \textbf{locally dense and globally sparse} networks \cite[Claim 5]{hack2025nib}, which are precisely the networks where we expect the KCN and NIB methods to be accurate. In general, the equations in the unbounded KCN and NIB methods may be different, and part of the extra complexity in the KCN may be used to compute some correlations more precisely. 

\section{Matrix spectrum via the NIB-method}
\label{sec: matrix spec}

We explain how to deal with matrix spectra through the NIB-method in this section, distinguishing the cases where the associated graph $\mg_{\mat{A}}$ fulfills the loop bound (Section \ref{sec: matrix bounded})
 from those where it does not (Section \ref{sec: matrix unbounded}).
Regarding the discussion in Section \ref{sec: single rv factors}, we will show that self-loops cannot be avoided in this application (Section \ref{sec: matrix self loops}).

\subsection{r-bounded loops}
\label{sec: matrix bounded}

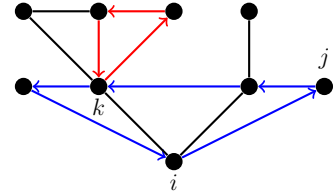
\begin{figure}[htb]
 \centering
\begin{tikzpicture}[scale=0.5, every node/.style={transform shape}]
\node[small dot, label={[anchor=east,below=4.5mm, thick, font=\fontsize{18}{18}\selectfont, thick]90:\textbf{$i$}}] at (0,0) (D) {};

\node[small dot] at (-4,2) (M1) {};
\node[small dot,  label={[anchor=east,below=4.5mm, thick, font=\fontsize{18}{18}\selectfont, thick]90:\textbf{$k$}}] at (-2,2) (M2) {};
\node[small dot] at (2,2) (M3) {};
\node[small dot, label={[anchor=east,above=1mm, thick, font=\fontsize{18}{18}\selectfont, thick]90:\textbf{$j$}}] at (4,2) (M4) {};

\node[small dot] at (-4,4) (U1) {};
\node[small dot] at (-2,4) (U2) {};
\node[small dot] at (0,4) (U3) {};
\node[small dot] at (2,4) (U4) {};

\path[draw,thick,->,blue]
    (D) edge node {} (M4)
 (M1) edge node {} (D)
  (M4) edge node {} (M3)
   (M3) edge node {} (M2)
    (M2) edge node {} (M1)
    ;
    
    \path[draw,thick,->,red]
    (U2) edge node {} (M2)
     (U3) edge node {} (U2)
 (M2) edge node {} (U3)
    ;

\path[draw,thick,-]
    (D) edge node {} (M2)
     (D) edge node {} (M3)
 (U1) edge node {} (M2)
  (U1) edge node {} (U2)
   (U4) edge node {} (M3)
    ;
\end{tikzpicture}
  \caption{Decomposition of an $i$-excursion in an $i$-excursion within the equivalence class $\overline{i \cap j}$ (in blue) together with a closed walks outside $\overline{i \cap j}$ that starts at $k \in (w_{\overline{i \cap j}} \cap \overline{i \cap j}) \smin \{i\}$ (in red).}
  \label{fig: excursion}
\end{figure}

If the loop bound is fulfilled, then any $i$-excursion can be decomposed as an $i$-excursion~$w_i$ within some equivalence class $w_i \subseteq \overline{i \cap j}$ together with some number of additional closed walks outside $\overline{i \cap j}$ that each start at some node $k \in (w_i \cap \overline{i \cap j}) \smin \{i\}$ and return some time later to $k$. Since the loop bound is fulfilled, the additional walks must return to the same node they started at. We give an example of such an excursion in Figure 
\ref{fig: excursion}.

To fix some notation, we assume that the length of the $i$-excursion~$w_i$ is~$l+1$, that is, $w_i$ visits $\ell$ (not necessarily distinct) nodes~$j_1,\dots, j_\ell \in \overline{i \cap j} \smin \{i\}$ within the equivalence class other than the starting node~$i$. Moreover, we take $s_{\overline{j \cap k} \to j}$ to be the length of some closed walk (if it exists) that starts and ends at~$j \in w_i \smin \{i\}$ and does not traverse any edges in some equivalence class containing $j$ that is different from $\overline{j \cap k} $. If no such a walk exists, then we take $s_{\overline{j \cap k} \to j}$ to be zero.

The total length of a \textbf{non-trivial} $i$-excursion $w_i$ \footnote{By non-trivial we mean $w_i \not \subseteq \overline{i \cap i}$, since otherwise the length is one by definition.} will be
\begin{equation*}
\ell+1+\textstyle{\sum_{j \in w_i \smin \{i\}, k: \overline{j \cap k} \in \equi{\cap} \smin \{\overline{i \cap j}\}} s_{\overline{j \cap k} \to j}},    
\end{equation*}
and the sum of the weights of all excursions of length~$s$ with $w_i$ as their foundation will be
\begin{widetext}
\begin{equation}
|w_i|\!\!\sum_{ \{ s_{\overline{j \cap k} \to j} : j \in w_i \smin \{i\}, k: \overline{j \cap k} \in \equi{\cap} \smin \{\overline{i \cap j}\} }
\displaystyle \prod_{j \in w_i \smin \{i\}} \prod_{k: \overline{j \cap k} \in \equi{\cap} \smin \{\overline{i \cap j}\}} X^{s_{\overline{j \cap k} \to j}}_{\overline{j \cap k} \to j} \!\! \text{ } \delta\bigl(s,\ell+1+\textstyle{\sum_{j \in w_i \smin \{i\}, k: \overline{j \cap k} \in \equi{\cap} \smin \{\overline{i \cap j}\}} s_{\overline{j \cap k} \to j}}\bigr),
\label{eq:excursions_length_s}
\end{equation}
\end{widetext}
where $|w_i|$ is the weight of $w_i$, and $X^{s}_{\overline{j \cap k} \to j}$ is the sum of weights of length-$s$ $j$-walks if the equivalence classes different from $\overline{j \cap k}$ that $j$ belongs to are removed from the graph. In fact, following the derivation of Eq.~\eqref{eq:X_i^s}, we have
\begin{equation}
X^{s}_{\overline{j \cap k} \to j} = \sum_{m=0}^{\infty} \left[ \sum_{s_1=1}^{\infty} \dots \sum_{s_m=1}^{\infty}  \delta \bigl(s, {\textstyle\sum_{u=1}^m} s_u \bigr) \prod_{u=1}^{m} Y^{s_u}_{\overline{j \cap k} \to j} \right],
\label{eq:X_ij^s}
\end{equation}
where, after the removal of the the equivalence classes different from $\overline{j \cap k}$ that $j$ belongs to,
the sum of weights of length-$s$ $j$-excursions is denoted by $Y^{s}_{\overline{j \cap k} \to j}$.

The last step before providing an equation for $H_i(z)$ is to find some computation leading to $Y_i^s$ in Eq.~\eqref{eq:closed_walk_gen_fun}. In fact, $Y_i^s$ can be decomposed as follows:
\begin{widetext}
\begin{equation}
\label{eq:Y_i^s}
\begin{split}
Y_i^s &= [\mat{A}]_{i i} \delta(s,1) +  \sum_{j \in \equi{(N_i \smin \{i\})}} \sum_{\ell_{\overline{i \cap j}}=0}^\infty \sum_{w_{\overline{i \cap j}} \in W_{\overline{i \cap j}}^{\ell_{\overline{i \cap j}}}} |w_{\overline{i \cap j}}| \sum_{\{ s_{\overline{k \cap q} \to k} : k \in w_{\overline{i \cap j}} \smin \{i\}, q: \overline{k \cap q} \in \equi{\cap} \smin \{\overline{i \cap j}\} \}} \displaystyle \prod_{k \in w_{\overline{i \cap j}} \smin \{i\}} \prod_{q: \overline{k \cap q} \in \equi{\cap} \smin \{\overline{i \cap j}\}} X^{s_{\overline{k \cap q} \to k}}_{\overline{k \cap q} \to k} \\
&\times \delta\bigl(s,\ell_{\overline{i \cap j}}+1+\textstyle{\sum_{k \in w_{\overline{i \cap j}} \smin \{i\}, q \in \equi{(N_k \smin \{i,j\})}} s_{\overline{k \cap q} \to k} }\bigr), 
\end{split}
\end{equation}
where~$W_{\overline{i \cap j}}^{\ell_{\overline{i\cap j}}}$ is the set of $i$-excursions of length $\ell_{\overline{i\cap j}}+1$ when, except for $\overline{i\cap j}$, the edges in all equivalence classes that $i$ belongs to are removed.
By putting together Eqs.~\eqref{eq:closed_walk_gen_fun},~\eqref{eq:X_ij^s} and~\eqref{eq:Y_i^s}, we obtain
\begin{equation}
     \label{eq:Hi}
     \begin{split}
         H_i(z) &= [\mat{A}]_{i i} + \sum_{j \in \equi{(N_i \smin \{i\})}} \sum_{\ell_{\overline{i \cap j}}=0}^{\infty} \frac{1}{z^{\ell_{\overline{i \cap j}}}}\!\!\sum_{w_{\overline{i \cap j}} \in W_{\overline{i \cap j}}^{\ell_{\overline{i \cap j}}}} |w_{\overline{i \cap j}}| \prod_{k \in w_{\overline{i \cap j}}\smin \{i\}}\,\prod_{q: \overline{k \cap q} \in \equi{\cap} \smin \{\overline{i \cap j}\}} \sum_{m=0}^{\infty}\,\prod_{x=1}^{m}\,\sum_{s=1}^{\infty} \frac{Y^{s}_{\overline{k \cap q} \to k}}{z^{s}}\\
    &= [\mat{A}]_{i i} + \sum_{j \in \equi{(N_i \smin \{i\})}} \sum_{w_{\overline{i \cap j}} \in W_{\overline{i \cap j}}} \!|w_{\overline{i \cap j}}| \prod_{k \in w_{\overline{i \cap j}}\smin \{i\}} \prod_{q: \overline{k \cap q} \in \equi{\cap} \smin \{\overline{i \cap j}\}} \frac{1}{z - H_{\overline{k \cap q} \to k}(z)},
     \end{split}
\end{equation}
where $W_{\overline{i \cap j}} \equiv \bigcup_{\ell_{\overline{i\cap j}}} W_{\overline{i \cap j}}^{\ell_{\overline{i\cap j}}}$, and we have defined
\begin{equation}
H_{\overline{k \cap q} \to k}(z) \equiv \sum_{s=1}^{\infty} \frac{Y^{s}_{\overline{k \cap q} \to k}}{z^{s-1}}.
\end{equation}
In the same vein, we can derive
\begin{equation}
\label{eq:spectral_message2}
H_{\overline{k \cap q} \to k}(z) = 
\begin{cases}
	\sum_{w_{\overline{k \cap q}} \in W_{\overline{k \cap q}}} \!|w_{\overline{k \cap q}}| \prod_{s \in w_{\overline{k \cap q}} \smin \{k \}} \prod_{v: \overline{s \cap v} \in \equi{\cap} \smin \{\overline{k \cap q}\} } \frac{1}{z - H_{\overline{s \cap v} \to s}(z)} & \quad\mbox{if $k \neq q$,} \\
  [\mat{A}]_{k k} & \quad\mbox{if $k = q$.}
\end{cases}
\end{equation}
\end{widetext}

Equation~\eqref{eq:spectral_message2} is our message passing scheme for the spectral density: We begin with suitable starting values and then iterate these equations to convergence. Once converged, we infer $H_i(z)$ through Eq.~\eqref{eq:Hi} and then use it to compute the spectral density via \eqref{eq:final_rho}.   

In order to make this approach practical, we ought to have some efficient method to evaluate the sum in Eq.~\eqref{eq:spectral_message2}. Since this can be done along the lines of the KCN-method \cite[Supplementary Material]{cantwell2019message}, we simply state how to do it without entering into the details.

We begin by considering
\begin{equation*}
\vec{v}_{\overline{k \cap q} \to k, v} \equiv \biggl\lbrace\begin{array}{ll}
	[\mat{A}]_{k v} & \quad\mbox{if $(k,v) \in \overline{k \cap q}$,} \\
  0 & \quad\mbox{otherwise,}
\end{array}
\end{equation*}
the vector of matrix elements associated to edges connected to $k$ in $\overline{k \cap q}$,
and by defining
$\mathbf{A}^{\overline{k \cap q}}$ the adjacency matrix of the neighborhood of $\overline{k \cap q}$:
\begin{equation}
\left[\mat{A}^{\overline{k \cap q}}\right]_{s v} \equiv \biggl\lbrace\begin{array}{ll}
	[\mat{A}]_{s v} & \quad\mbox{if $s,v\neq k$ and $(s,v)\in \overline{k \cap q},$} \\
  0 & \quad\mbox{otherwise.}
\end{array}
\end{equation}
Lastly, we let $\mathbf{D}^{\overline{k \cap q} \to k}(z)$ be the diagonal matrix with entries 
\begin{equation*}
\left[D^{\overline{k \cap q} \to k}(z)\right]_{ss} \equiv \prod_{\overline{s \cap v} \neq \overline{k \cap q}} \left(z-H_{\overline{s \cap v} \to s}(z) \right),
\end{equation*}
and we obtain that, for $k \neq q$, Equation~\eqref{eq:spectral_message2} can then be written as
\begin{equation}
H_{\overline{k \cap q} \to k}(z) = \vec{v}_{\overline{k \cap q} \to k}^T \bigl( \mathbf{D}^{\overline{k \cap q} \to k}(z) - \mathbf{A}^{\overline{k \cap q}} \bigr)^{-1} \vec{v}_{\overline{k \cap q} \to k}.
\label{eq:matrix_method}
\end{equation}

Since the loop bound is fulfilled, the equations in this section provide exact results. Moreover, they provide an advantage regarding time complexity compared to the matrix spectra version of the KCN-method \cite{cantwell2019message}: Instead of inverting a matrix of dimension $|N_{j \smin i}| \times |N_{j \smin i}|$ to compute $H_{j \to i}(z)$ in the analogous of \eqref{eq:matrix_method} (or even of dimension $|N_{i}| \times |N_{i}|$ to compute $H_{i}(z)$), we invert one of size $|N_{i \cap j}| \times |N_{i \cap j}|$. (Recall that the complexity of matrix inversion is cubic in the size of its dimensions.)
In fact, following \cite[Claim 4]{hack2025nib}, we can again show that the approach to matrix spectra in this section is optimal in terms of time complexity.

\subsection{r-unbounded loops}
\label{sec: matrix unbounded}

If the loop bound is not fulfilled, then we can again use the maps defined in Section \ref{sec: nib unbounded} to extend our approach to matrix spectra from the bounded case in the spirit of the extension of the NIB-method to the unbounded case \cite{hack2025nib}.

Regarding message passing, and aside from the trivial messages $H_{k \cap k \to i \cap j}(z) = [\mat{A}]_{kk}$ for all $z$, the analogous of \eqref{eq:matrix_method} is
\begin{equation*}
\begin{split}
    H_{k \cap q \to i \cap j}(z) &\equiv (\vec{v}^{\overline{\pp_{i \cap j}}(N_{k \cap q})}_{k \cap q \to i \cap j})^T \bigl( \mathbf{D}^{k \cap q \to i \cap j} - \mathbf{A}^{k \cap q \to i \cap j}_{\overline{\pp_{i \cap j}}(N_{k \cap q})} \bigr)^{-1} \\
    &\times \vec{v}^{\overline{\pp_{i \cap j}}(N_{k \cap q})}_{k \cap q \to i \cap j},
\end{split}
\end{equation*}
where 
\begin{equation*}
\begin{split}
\vec{v}^{\overline{\pp_{i \cap j}}(N_{k \cap q})}_{k \cap q \to i \cap j, v} \equiv \biggl\lbrace\begin{array}{ll}
	[\mat{A}]_{k v} & \quad\mbox{if $(k,v) \in N_{k \cap q} \smin \overline{\pp_{i \cap j}}(N_{k \cap q})$,} \\
  0 & \quad\mbox{otherwise;}
\end{array}
\end{split}
\end{equation*}
\begin{equation*}
\left[\mat{A}^{k \cap q \to i \cap j}_{\overline{\pp_{i \cap j}}(N_{k \cap q})}\right]_{s v} \equiv \biggl\lbrace\begin{array}{ll}
	[\mat{A}]_{s v} & \quad\mbox{if $s,v\neq k$ and} \\
	& \quad\mbox{$(s,v)\in N_{k \cap q} \smin \overline{\pp_{i \cap j}}(N_{k \cap q})$,} \\
  0 & \quad\mbox{otherwise.}
\end{array}
\end{equation*}
Moreover, $\mathbf{D}^{k \cap q \to i \cap j}(z)$ is the diagonal matrix with entries 
\begin{equation*}
\left[\mat{D}^{k \cap q \to i \cap j} (z)\right]_{ss} \equiv \prod_{v \in N_s} \left(z-H_{s \cap v \to k \cap q}(z) \right).
\end{equation*}

To conclude, the inference formula \eqref{eq:Hi} for $H_i$ becomes
\begin{equation*}
\begin{split}
    H_i(z) &= [\mat{A}]_{ii} + \sum_{j \in N_i \smin \{i\}} \sum_{w_{i \cap j} \in W_{i \cap j \smin \overline{\ppq_{i}}(N_{i \cap j})}} \!|w_{i \cap j}| \times  \\
    &\prod_{k \in w_{i \cap j}\smin \{i\}} \prod_{q \in N_k: N_{k \cap q} \smin \overline{\pp_{i \cap j}}(N_{k \cap q}) \neq \emptyset} \frac{1}{z - H_{k \cap q \to i \cap j}(z)}\,,
    %\label{eq:Hi}
\end{split}
\end{equation*}
where $W_{i \cap j \smin \overline{\ppq_{i}}(N_{i \cap j})}$ is the set of $i$-excursions that use edges within  $N_{i \cap j} \smin \overline{\ppq_{i}}(N_{i \cap j})$.

Since the loop bound is not fulfilled, the equations only provide approximate results. The time complexity advantage compared to the KCN-method remains provided we consider locally dense and globally sparse networks \cite[Claim 5]{hack2025nib}, which are precisely the graphs where we expect the KCN and NIB methods to be accurate. In general, the equations in the unbounded KCN and NIB methods may be different, and part of the extra complexity in the KCN may be used to compute some correlations more precisely.

\subsection{Self-loops}
\label{sec: matrix self loops}

Let us return to the discussion in Section \ref{sec: single rv factors} and assume we would like to use the original version of the NIB-method to compute matrix spectra. To do so, we take some enumeration $(k_i)_{i=1}^m \subseteq \mv$ of the edges with self-loops, and consider the sequence of weighted graphs $(\mg^i_{\mat{A}}=(\mv^i,\me^i,\mw^i))_{i=0}^m$ defined recursively as follows:
\begin{itemize}
    \item $\mg^0_{\mat{A}} \equiv \mg_{\mat{A}}$.
    \item For $1 \leq i \leq m$, $\mv^{i}=\mv^{i-1}$, $\me^{i}=\me^{i-1} \smin \{(k_i,k_i)\}$, and, taking some $j_{k_i} \in \mv'$ with $j_{k_i} \neq k_i$,
    \begin{equation*}
w^{i}_e \equiv 
\begin{cases}
	w^{i-1}_e \cdot [\mat{A}]_{k_ik_i} & \quad\mbox{if $e = (j_{k_i},k_i)$,} \\
  w^{i-1}_e & \quad\mbox{if $e \neq (j_{k_i},k_i)$.}
\end{cases}
\end{equation*}
    for all $e \in \me^{i}$.
\end{itemize}
The final graph $\mg^m_{\mat{A}}$ corresponds to the original version of the NIB-method applied to matrix spectra. However, the excursions consisting of a single self-loop are not recoverable in $\mg^m_{\mat{A}}$. This implies several issues from \eqref{eq:excursions_length_s} onwards.

\section{Conclusion}

We have extended the NIB-method to percolation and the computation of matrix spectra, showing that one can also achieve an improvement on the KCN-method in these applications. If either the loop bound is fulfilled or it is not fulfilled but the graph is locally dense and globally sparse, then the improvement can be shown analytically, as we have argued. If the loop bound is not fulfilled, then it is reasonable to assume that the numerical evidence comparing the performance of the KCN and NIB methods in the context of probabilistic graphical models will extend to the applications discussed here. However, providing such numerical evidence remains a task for the future. 

Regarding the comparison to previous literature other than the KCN-method, we can make the following remarks:
\begin{itemize}
    \item Concerning percolation, it was already argued  \cite{cantwell2019message} that the KCN-method and classical direct simulations compute different quantities. That is, while the latter only considers a single realized graph and one would need to perform several runs in order to obtain average values, the former directly provides averaged values. The NIB-method also computes averaged values as well. After the introduction of the KCN-method, a motif-based message passing approach \cite{mann2023belief} was developed. Although it was conceived for a different purpose, it is important to note that its message passing algorithm is limited to some specific graphs and it requires some graph-dependent analytical derivations. This contrasts with the generality of the KCN and NIB methods. 
    
    \item Concerning matrix spectra, the KCN-method can substantially outperform traditional methods \cite{cantwell2019message}, thus enabling the computation of the spectra of some previously inaccessible large systems. The NIB-method can extend the set of accessible systems even further.
\end{itemize}

As future research directions, let us emphasize the following:
\begin{itemize}
    \item In the context of inference, it would be important to extend the NIB-method from networks to general graphical models. 
    \item It would be interesting to extend the KCN and NIB methods to other applications, like epidemic models or graph coloring. A very interesting use case could be the computation of thresholds in the context of quantum error correction and the erasure channel \cite{stace2009thresholds,delfosse2010quantum}. This is closely related to percolation and it has practical relevance since it addresses a simplified error model that has proven to be key in order to gain insight regarding decoding.
    \item The application of these methods to compute the spectra of non-symmetric matrices has not been developed yet, and it seems like one would need to make fundamental modifications to the the symmetric case. 
\end{itemize}

\bibliography{__main}

\begin{thebibliography}{23}%
\makeatletter
\providecommand \@ifxundefined [1]{%
 \@ifx{#1\undefined}
}%
\providecommand \@ifnum [1]{%
 \ifnum #1\expandafter \@firstoftwo
 \else \expandafter \@secondoftwo
 \fi
}%
\providecommand \@ifx [1]{%
 \ifx #1\expandafter \@firstoftwo
 \else \expandafter \@secondoftwo
 \fi
}%
\providecommand \natexlab [1]{#1}%
\providecommand \enquote  [1]{``#1''}%
\providecommand \bibnamefont  [1]{#1}%
\providecommand \bibfnamefont [1]{#1}%
\providecommand \citenamefont [1]{#1}%
\providecommand \href@noop [0]{\@secondoftwo}%
\providecommand \href [0]{\begingroup \@sanitize@url \@href}%
\providecommand \@href[1]{\@@startlink{#1}\@@href}%
\providecommand \@@href[1]{\endgroup#1\@@endlink}%
\providecommand \@sanitize@url [0]{\catcode `\\12\catcode `\$12\catcode `\&12\catcode `\#12\catcode `\^12\catcode `\_12\catcode `\%12\relax}%
\providecommand \@@startlink[1]{}%
\providecommand \@@endlink[0]{}%
\providecommand \url  [0]{\begingroup\@sanitize@url \@url }%
\providecommand \@url [1]{\endgroup\@href {#1}{\urlprefix }}%
\providecommand \urlprefix  [0]{URL }%
\providecommand \Eprint [0]{\href }%
\providecommand \doibase [0]{https://doi.org/}%
\providecommand \selectlanguage [0]{\@gobble}%
\providecommand \bibinfo  [0]{\@secondoftwo}%
\providecommand \bibfield  [0]{\@secondoftwo}%
\providecommand \translation [1]{[#1]}%
\providecommand \BibitemOpen [0]{}%
\providecommand \bibitemStop [0]{}%
\providecommand \bibitemNoStop [0]{.\EOS\space}%
\providecommand \EOS [0]{\spacefactor3000\relax}%
\providecommand \BibitemShut  [1]{\csname bibitem#1\endcsname}%
\let\auto@bib@innerbib\@empty
%</preamble>
\bibitem [{\citenamefont {Richardson}\ and\ \citenamefont {Urbanke}(2008)}]{richardson2008}%
  \BibitemOpen
  \bibfield  {author} {\bibinfo {author} {\bibfnamefont {T.}~\bibnamefont {Richardson}}\ and\ \bibinfo {author} {\bibfnamefont {R.}~\bibnamefont {Urbanke}},\ }\href@noop {} {\emph {\bibinfo {title} {Modern coding theory}}}\ (\bibinfo  {publisher} {Cambridge university press},\ \bibinfo {year} {2008})\BibitemShut {NoStop}%
\bibitem [{\citenamefont {Mezard}\ and\ \citenamefont {Montanari}(2009)}]{mezard2009}%
  \BibitemOpen
  \bibfield  {author} {\bibinfo {author} {\bibfnamefont {M.}~\bibnamefont {Mezard}}\ and\ \bibinfo {author} {\bibfnamefont {A.}~\bibnamefont {Montanari}},\ }\href@noop {} {\emph {\bibinfo {title} {Information, physics, and computation}}}\ (\bibinfo  {publisher} {Oxford University Press},\ \bibinfo {year} {2009})\BibitemShut {NoStop}%
\bibitem [{\citenamefont {Liu}\ and\ \citenamefont {Poulin}(2019)}]{liu2019}%
  \BibitemOpen
  \bibfield  {author} {\bibinfo {author} {\bibfnamefont {Y.-H.}\ \bibnamefont {Liu}}\ and\ \bibinfo {author} {\bibfnamefont {D.}~\bibnamefont {Poulin}},\ }\bibfield  {title} {\bibinfo {title} {Neural belief-propagation decoders for quantum error-correcting codes},\ }\href@noop {} {\bibfield  {journal} {\bibinfo  {journal} {Physical review letters}\ }\textbf {\bibinfo {volume} {122}},\ \bibinfo {pages} {200501} (\bibinfo {year} {2019})}\BibitemShut {NoStop}%
\bibitem [{\citenamefont {Yedidia}\ \emph {et~al.}(2000)\citenamefont {Yedidia}, \citenamefont {Freeman},\ and\ \citenamefont {Weiss}}]{yedidia2000generalized}%
  \BibitemOpen
  \bibfield  {author} {\bibinfo {author} {\bibfnamefont {J.~S.}\ \bibnamefont {Yedidia}}, \bibinfo {author} {\bibfnamefont {W.}~\bibnamefont {Freeman}},\ and\ \bibinfo {author} {\bibfnamefont {Y.}~\bibnamefont {Weiss}},\ }\bibfield  {title} {\bibinfo {title} {Generalized belief propagation},\ }\href@noop {} {\bibfield  {journal} {\bibinfo  {journal} {Advances in neural information processing systems}\ }\textbf {\bibinfo {volume} {13}} (\bibinfo {year} {2000})}\BibitemShut {NoStop}%
\bibitem [{\citenamefont {Welling}(2004)}]{welling2004choice}%
  \BibitemOpen
  \bibfield  {author} {\bibinfo {author} {\bibfnamefont {M.}~\bibnamefont {Welling}},\ }\bibfield  {title} {\bibinfo {title} {On the choice of regions for generalized belief propagation},\ }in\ \href@noop {} {\emph {\bibinfo {booktitle} {Proceedings of the 20th Conference on Uncertainty in Artificial Intelligence}}}\ (\bibinfo {organization} {AUAI Press},\ \bibinfo {year} {2004})\ pp.\ \bibinfo {pages} {585--592}\BibitemShut {NoStop}%
\bibitem [{\citenamefont {Hack}\ \emph {et~al.}(2024)\citenamefont {Hack}, \citenamefont {Mendl},\ and\ \citenamefont {Paler}}]{hack2024belief}%
  \BibitemOpen
  \bibfield  {author} {\bibinfo {author} {\bibfnamefont {P.}~\bibnamefont {Hack}}, \bibinfo {author} {\bibfnamefont {C.~B.}\ \bibnamefont {Mendl}},\ and\ \bibinfo {author} {\bibfnamefont {A.}~\bibnamefont {Paler}},\ }\bibfield  {title} {\bibinfo {title} {Belief propagation for general graphical models with loops},\ }\href@noop {} {\bibfield  {journal} {\bibinfo  {journal} {arXiv preprint arXiv:2411.04957}\ } (\bibinfo {year} {2024})}\BibitemShut {NoStop}%
\bibitem [{\citenamefont {Hack}(2025)}]{hack2025nib}%
  \BibitemOpen
  \bibfield  {author} {\bibinfo {author} {\bibfnamefont {P.}~\bibnamefont {Hack}},\ }\bibfield  {title} {\bibinfo {title} {Belief propagation for networks with loops: The neighborhoods-intersections-based method},\ }\href@noop {} {\bibfield  {journal} {\bibinfo  {journal} {arXiv preprint arXiv:2506.13791}\ } (\bibinfo {year} {2025})}\BibitemShut {NoStop}%
\bibitem [{\citenamefont {Cantwell}\ and\ \citenamefont {Newman}(2019)}]{cantwell2019message}%
  \BibitemOpen
  \bibfield  {author} {\bibinfo {author} {\bibfnamefont {G.~T.}\ \bibnamefont {Cantwell}}\ and\ \bibinfo {author} {\bibfnamefont {M.~E.}\ \bibnamefont {Newman}},\ }\bibfield  {title} {\bibinfo {title} {Message passing on networks with loops},\ }\href@noop {} {\bibfield  {journal} {\bibinfo  {journal} {Proceedings of the National Academy of Sciences}\ }\textbf {\bibinfo {volume} {116}},\ \bibinfo {pages} {23398} (\bibinfo {year} {2019})}\BibitemShut {NoStop}%
\bibitem [{\citenamefont {Stauffer}\ and\ \citenamefont {Aharony}(2018)}]{stauffer2018introduction}%
  \BibitemOpen
  \bibfield  {author} {\bibinfo {author} {\bibfnamefont {D.}~\bibnamefont {Stauffer}}\ and\ \bibinfo {author} {\bibfnamefont {A.}~\bibnamefont {Aharony}},\ }\href@noop {} {\emph {\bibinfo {title} {Introduction to percolation theory}}}\ (\bibinfo  {publisher} {Taylor \& Francis},\ \bibinfo {year} {2018})\BibitemShut {NoStop}%
\bibitem [{\citenamefont {Karrer}\ \emph {et~al.}(2014)\citenamefont {Karrer}, \citenamefont {Newman},\ and\ \citenamefont {Zdeborov{\'a}}}]{karrer2014percolation}%
  \BibitemOpen
  \bibfield  {author} {\bibinfo {author} {\bibfnamefont {B.}~\bibnamefont {Karrer}}, \bibinfo {author} {\bibfnamefont {M.~E.}\ \bibnamefont {Newman}},\ and\ \bibinfo {author} {\bibfnamefont {L.}~\bibnamefont {Zdeborov{\'a}}},\ }\bibfield  {title} {\bibinfo {title} {Percolation on sparse networks},\ }\href@noop {} {\bibfield  {journal} {\bibinfo  {journal} {Physical review letters}\ }\textbf {\bibinfo {volume} {113}},\ \bibinfo {pages} {208702} (\bibinfo {year} {2014})}\BibitemShut {NoStop}%
\bibitem [{\citenamefont {Newman}\ and\ \citenamefont {Ziff}(2000)}]{newman2000efficient}%
  \BibitemOpen
  \bibfield  {author} {\bibinfo {author} {\bibfnamefont {M.}~\bibnamefont {Newman}}\ and\ \bibinfo {author} {\bibfnamefont {R.~M.}\ \bibnamefont {Ziff}},\ }\bibfield  {title} {\bibinfo {title} {Efficient monte carlo algorithm and high-precision results for percolation},\ }\href@noop {} {\bibfield  {journal} {\bibinfo  {journal} {Physical Review Letters}\ }\textbf {\bibinfo {volume} {85}},\ \bibinfo {pages} {4104} (\bibinfo {year} {2000})}\BibitemShut {NoStop}%
\bibitem [{\citenamefont {Kirkley}\ \emph {et~al.}(2021)\citenamefont {Kirkley}, \citenamefont {Cantwell},\ and\ \citenamefont {Newman}}]{kirkley2021belief}%
  \BibitemOpen
  \bibfield  {author} {\bibinfo {author} {\bibfnamefont {A.}~\bibnamefont {Kirkley}}, \bibinfo {author} {\bibfnamefont {G.~T.}\ \bibnamefont {Cantwell}},\ and\ \bibinfo {author} {\bibfnamefont {M.}~\bibnamefont {Newman}},\ }\bibfield  {title} {\bibinfo {title} {Belief propagation for networks with loops},\ }\href@noop {} {\bibfield  {journal} {\bibinfo  {journal} {Science Advances}\ }\textbf {\bibinfo {volume} {7}},\ \bibinfo {pages} {eabf1211} (\bibinfo {year} {2021})}\BibitemShut {NoStop}%
\bibitem [{\citenamefont {Bianconi}\ and\ \citenamefont {Dorogovtsev}(2024)}]{bianconi2024theory}%
  \BibitemOpen
  \bibfield  {author} {\bibinfo {author} {\bibfnamefont {G.}~\bibnamefont {Bianconi}}\ and\ \bibinfo {author} {\bibfnamefont {S.~N.}\ \bibnamefont {Dorogovtsev}},\ }\bibfield  {title} {\bibinfo {title} {Theory of percolation on hypergraphs},\ }\href@noop {} {\bibfield  {journal} {\bibinfo  {journal} {Physical Review E}\ }\textbf {\bibinfo {volume} {109}},\ \bibinfo {pages} {014306} (\bibinfo {year} {2024})}\BibitemShut {NoStop}%
\bibitem [{\citenamefont {Xiong}\ \emph {et~al.}(2025)\citenamefont {Xiong}, \citenamefont {Dong}, \citenamefont {Liu}, \citenamefont {Zhou},\ and\ \citenamefont {Liu}}]{xiong2025regulation}%
  \BibitemOpen
  \bibfield  {author} {\bibinfo {author} {\bibfnamefont {K.}~\bibnamefont {Xiong}}, \bibinfo {author} {\bibfnamefont {H.}~\bibnamefont {Dong}}, \bibinfo {author} {\bibfnamefont {Y.}~\bibnamefont {Liu}}, \bibinfo {author} {\bibfnamefont {M.}~\bibnamefont {Zhou}},\ and\ \bibinfo {author} {\bibfnamefont {W.}~\bibnamefont {Liu}},\ }\bibfield  {title} {\bibinfo {title} {Regulation of thermal transport by cycle structures in complex networks},\ }\href@noop {} {\bibfield  {journal} {\bibinfo  {journal} {Chaos, Solitons \& Fractals}\ }\textbf {\bibinfo {volume} {191}},\ \bibinfo {pages} {115766} (\bibinfo {year} {2025})}\BibitemShut {NoStop}%
\bibitem [{\citenamefont {Castro~Guzman}\ \emph {et~al.}(2025)\citenamefont {Castro~Guzman}, \citenamefont {Stadler},\ and\ \citenamefont {Fujita}}]{castro2025message}%
  \BibitemOpen
  \bibfield  {author} {\bibinfo {author} {\bibfnamefont {G.~E.}\ \bibnamefont {Castro~Guzman}}, \bibinfo {author} {\bibfnamefont {P.~F.}\ \bibnamefont {Stadler}},\ and\ \bibinfo {author} {\bibfnamefont {A.}~\bibnamefont {Fujita}},\ }\bibfield  {title} {\bibinfo {title} {A message-passing approach to obtain the trace of matrix functions with applications to network analysis},\ }\href@noop {} {\bibfield  {journal} {\bibinfo  {journal} {Numerical Algorithms}\ ,\ \bibinfo {pages} {1}} (\bibinfo {year} {2025})}\BibitemShut {NoStop}%
\bibitem [{\citenamefont {Newman}(2023)}]{newman2023message}%
  \BibitemOpen
  \bibfield  {author} {\bibinfo {author} {\bibfnamefont {M.}~\bibnamefont {Newman}},\ }\bibfield  {title} {\bibinfo {title} {Message passing methods on complex networks},\ }\href@noop {} {\bibfield  {journal} {\bibinfo  {journal} {Proceedings of the Royal Society A}\ }\textbf {\bibinfo {volume} {479}},\ \bibinfo {pages} {20220774} (\bibinfo {year} {2023})}\BibitemShut {NoStop}%
\bibitem [{\citenamefont {Nadakuditi}\ and\ \citenamefont {Newman}(2013)}]{nadakuditi2013spectra}%
  \BibitemOpen
  \bibfield  {author} {\bibinfo {author} {\bibfnamefont {R.~R.}\ \bibnamefont {Nadakuditi}}\ and\ \bibinfo {author} {\bibfnamefont {M.~E.}\ \bibnamefont {Newman}},\ }\bibfield  {title} {\bibinfo {title} {Spectra of random graphs with arbitrary expected degrees},\ }\href@noop {} {\bibfield  {journal} {\bibinfo  {journal} {Physical Review E—Statistical, Nonlinear, and Soft Matter Physics}\ }\textbf {\bibinfo {volume} {87}},\ \bibinfo {pages} {012803} (\bibinfo {year} {2013})}\BibitemShut {NoStop}%
\bibitem [{Note1()}]{Note1}%
  \BibitemOpen
  \bibinfo {note} {More specifically, given some $x \in \protect \mathbb R$ of interest, we use $z= x + i \eta _0$ for some fixed $\eta _0>0$. For instance, in \cite {cantwell2019message}, $\eta _0=0.05, 0.01$. When using the message passing methods that we will present later on, one runs them with such a fixed value and, when convergence is achieved, we simply take the imaginary part of $\rho (z)$. Hence, we will run the algorithm once for each value of $x$.}\BibitemShut {Stop}%
\bibitem [{Note2()}]{Note2}%
  \BibitemOpen
  \bibinfo {note} {We could avoid this condition and simply take $\protect \mathcal G_A$ to have full connectivity with some weights being null. However, since our message passing methods will be exploiting the sparsity of $\protect \mathbf {A}$, it is more convenient to associate a sparse graph $\protect \mathcal G_{\protect \mathbf {A}}$ to a sparse matrix $\protect \mathbf {A}$.}\BibitemShut {Stop}%
\bibitem [{Note3()}]{Note3}%
  \BibitemOpen
  \bibinfo {note} {By non-trivial we mean $w_i \not \subseteq \protect \overline {i \cap i}$, since otherwise the length is one by definition.}\BibitemShut {Stop}%
\bibitem [{\citenamefont {Mann}\ and\ \citenamefont {Dobson}(2023)}]{mann2023belief}%
  \BibitemOpen
  \bibfield  {author} {\bibinfo {author} {\bibfnamefont {P.}~\bibnamefont {Mann}}\ and\ \bibinfo {author} {\bibfnamefont {S.}~\bibnamefont {Dobson}},\ }\bibfield  {title} {\bibinfo {title} {Belief propagation on networks with cliques and chordless cycles},\ }\href@noop {} {\bibfield  {journal} {\bibinfo  {journal} {Physical Review E}\ }\textbf {\bibinfo {volume} {107}},\ \bibinfo {pages} {054303} (\bibinfo {year} {2023})}\BibitemShut {NoStop}%
\bibitem [{\citenamefont {Stace}\ \emph {et~al.}(2009)\citenamefont {Stace}, \citenamefont {Barrett},\ and\ \citenamefont {Doherty}}]{stace2009thresholds}%
  \BibitemOpen
  \bibfield  {author} {\bibinfo {author} {\bibfnamefont {T.~M.}\ \bibnamefont {Stace}}, \bibinfo {author} {\bibfnamefont {S.~D.}\ \bibnamefont {Barrett}},\ and\ \bibinfo {author} {\bibfnamefont {A.~C.}\ \bibnamefont {Doherty}},\ }\bibfield  {title} {\bibinfo {title} {Thresholds for topological codes in the presence of loss},\ }\href@noop {} {\bibfield  {journal} {\bibinfo  {journal} {Physical review letters}\ }\textbf {\bibinfo {volume} {102}},\ \bibinfo {pages} {200501} (\bibinfo {year} {2009})}\BibitemShut {NoStop}%
\bibitem [{\citenamefont {Delfosse}\ and\ \citenamefont {Z{\'e}mor}(2010)}]{delfosse2010quantum}%
  \BibitemOpen
  \bibfield  {author} {\bibinfo {author} {\bibfnamefont {N.}~\bibnamefont {Delfosse}}\ and\ \bibinfo {author} {\bibfnamefont {G.}~\bibnamefont {Z{\'e}mor}},\ }\bibfield  {title} {\bibinfo {title} {Quantum erasure-correcting codes and percolation on regular tilings of the hyperbolic plane},\ }in\ \href@noop {} {\emph {\bibinfo {booktitle} {2010 IEEE Information Theory Workshop}}}\ (\bibinfo {organization} {IEEE},\ \bibinfo {year} {2010})\ pp.\ \bibinfo {pages} {1--5}\BibitemShut {NoStop}%
\end{thebibliography}%

\end{document}